\documentclass[epj]{svjour}
\usepackage{graphicx}
\input{epsf}
\usepackage{amssymb}
\begin{document}

\title{Relativistic graphene ratchet on semidisk Galton board}

\author{L.Ermann \and D.L.Shepelyansky}
\institute{Laboratoire de Physique Th\'eorique du CNRS, IRSAMC, 
Universit\'e de Toulouse, UPS, F-31062 Toulouse, France
\and
http://www.quantware.ups-tlse.fr}

\titlerunning{Relativistic graphene ratchet}
\authorrunning{L.Ermann and D.L.Shepelyansky}

\abstract{ Using extensive Monte Carlo simulations 
we study numerically and analytically
a photogalvanic effect, or ratchet,   of directed electron transport 
induced by a microwave radiation
on a semidisk Galton board of antidots in graphene. A comparison between usual 
two-dimensional electron gas (2DEG) and electrons 
in graphene shows that ratchet currents are comparable at very low
temperatures. However,   a large mean free path in graphene
should allow to have a strong ratchet transport at room temperatures.
Also in graphene the ratchet transport emerges
even for unpolarized radiation.
These properties open promising possibilities for room temperature
graphene based sensitive photogalvanic 
detectors of microwave and terahertz radiation.
}


\date{Dated: October 21, 2010; Revised: November 17, 2010 }

\maketitle


Graphene \cite{geim2004} is a new two-dimensional material 
with a variety of  fascinating physical properties (see e.g. \cite{geimkim}).
One of them is a relativistic dispersion
law for electron dynamics with an effective 
``light'' velocity $s \approx 10^8 cm/s$ \cite{geimrmp2009},
another is a high mobility at room temperature
which in suspended graphene reaches 
$\mu \approx 200,000 cm^2/Vs$ \cite{geimmobil,stormer2008,andrei2008}.
Our theoretical studies  predict that these properties lead to 
emergence of a strong photogalvanic effect
induced by radiation at room temperature in a semidisk antidot array
placed in graphene plane.
Such samples  can function as a new type of 
room temperature sensors of microwave and terahertz radiation.

\begin{figure}
\includegraphics[width=0.48\textwidth]{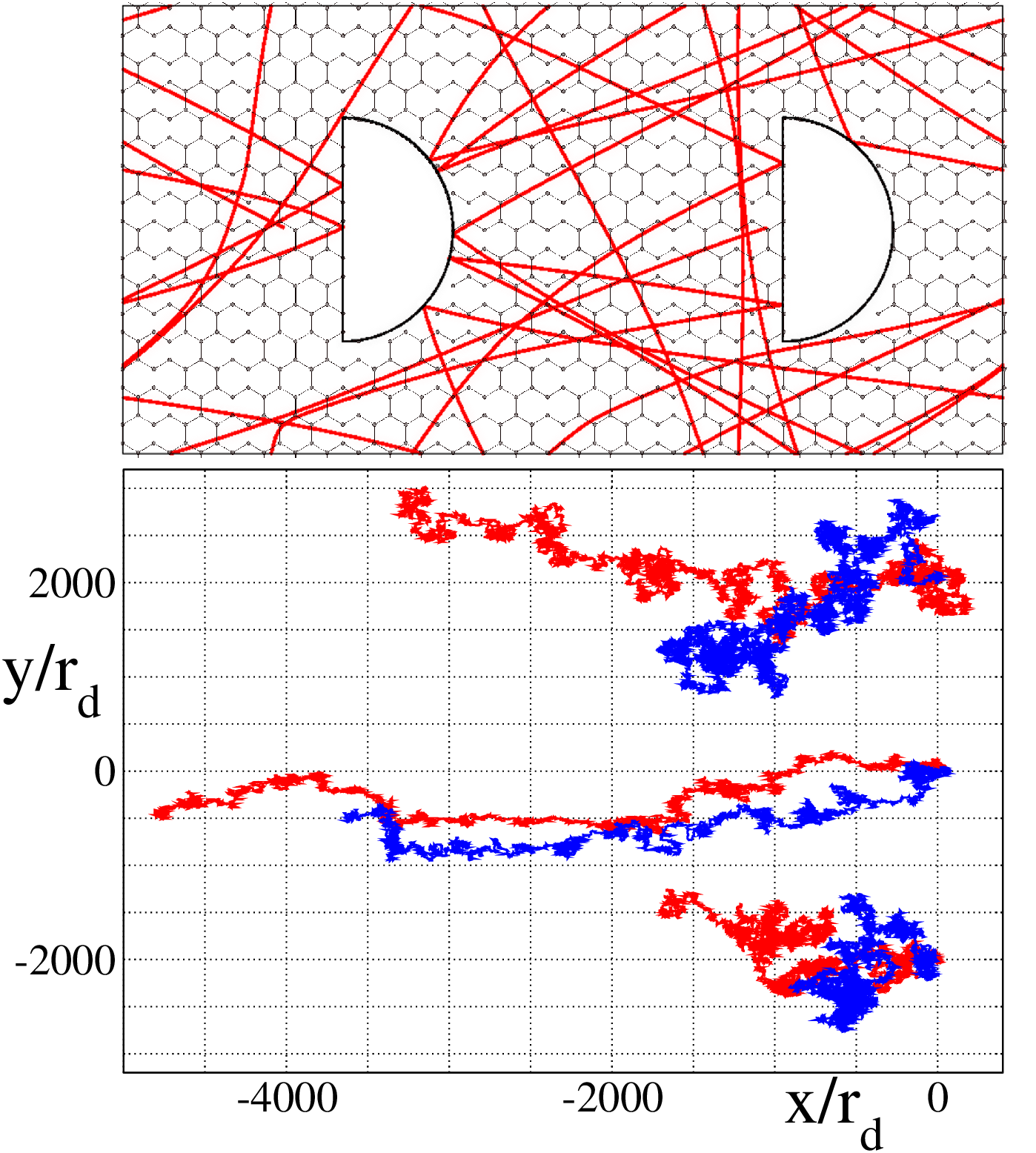}
\vglue -0.2cm
\caption{(Color online) Top panel shows  one electron 
trajectory on semidisk Galton square board (with periodic 
boundary conditions for two cells, no impurities, time
$t \approx 100 r_d/s$ and $f r_d/E_F=0.05$),
graphene structure is shown in a schematic way.
Bottom panel shows two ratchet trajectories
on a longer time for graphene 
(with initial conditions $(x,y)=(0,2000)$ and $t \approx 2\times10^5 
r_d/s$), model graphene 
(with $(x,y)=(0,0)$ and $t \approx 10^5 r_d/s$) and 2DEG
(with $(x/r_d,y/r_d)=(0,-2000)$ and  same $t$) 
all of them at the same $p_F$
and $T/E_F=0.02$; here 
$R/r_d=4$, $\theta=0$, $\omega r_d/s = 0.5$, $V_F/s=2$,
and $fr_d/E_f=0.1$ (red/gray curve),
$0.025$ (blue/black curve); impurity parameters are
$s \tau_i/r_d =5$, $\alpha_i=\pi/10$.
}
\label{fig1}
\end{figure}

Ratchet transport, induced in asymmetric systems 
by {\it ac-}driving with zero mean force,
attracted a significant interest of scientific community in view  of 
various biological applications of 
Brownian motors \cite{prost,hanggi,reimann}.
Experimental observations of electron ratchet transport 
in asymmetric antidot arrays in semiconductor heterostructures
have been reported in \cite{lorke,linke,song,portal2008}.
A detailed theory of ratchet transport of 2DEG on semidisk Galton 
board  was developed in \cite{alik2006,entin2007}
for noninteracting electrons, and it was shown that
the effect remains even in presence of strong interactions
\cite{entin2008}. The theoretical predictions on polarization dependence
have been confirmed in recent experiments \cite{portal2008}.
According to theory \cite{alik2006,entin2007}
the velocity of ratchet flow, induced by monochromatic linear polarized
microwave force $\mathbf{f} \cos{\omega t}$ with a frequency $\omega$,
has a polarization dependence:
\begin{eqnarray}
\label{eq1}
({\bar v}_x,{\bar v}_y)  =  
C_F V_F (f r_d/E_F)^2 (-\cos(2\theta), \sin(2\theta)) \; , 
\end{eqnarray}
where $\theta$ is an angle between the polarization direction and 
$x-$axis of semidisk array shown in Fig.~\ref{fig1}, 
$f$ is amplitude of microwave force, $V_F$, $E_F$ 
are Fermi velocity and energy,
and $C_F$ is a numerical factor depending 
on the ratio of periodic cell size
$R$ to semidisk radius $r_d$. 
In the limit of low density of semidisks with $R \gg r_d$ 
the theory \cite{entin2007}
gives $C_F \propto (\ell_s/r_d)^2/(1+(\omega \ell_s /V_F)^2)$,
with $\ell_s \sim R^2/r_d \gg \ell_i$,
where $\ell_i$ is the mean free path related to impurity scattering. 
The typical 
parameters of experiment \cite{portal2008}
are $f/e \sim 1 V/cm$, $r_d \sim 1 \mu m$, $R/r_d \approx 4$,
electron density $n_e \approx 2.5 \cdot 10^{11} cm^{-2}$
with $V_F \approx 2.2 \cdot 10^7 cm/s$, $E_ F \approx 100 K$.
For these conditions we have $f r_d/E_F \approx 1/100$, $C_F \approx 0.4$,
so that the velocity of ratchet flow remains
relatively small  but well visible experimentally.
Experimental data confirm the linear dependence of ratchet current 
on microwave power \cite{portal2008}.
The relation (\ref{eq1}) assumes that 
$\omega r_d/V_F <1$ and that
the mean free path $\ell_i$
remains larger   semidisk size.
For $\ell_i < r_d$ the array asymmetry disappears due to disorder
scattering and ratchet velocity goes to zero.

The theory (\ref{eq1}) assumes the usual quadratic dispersion
law for electron dynamics $E=p^2/2m$,
while graphene has a linear relativistic dispersion $E = s p$,
so that such a case requires a separate analysis. 
Recently rectification and photon drag  in graphene 
started to attract experimental  \cite{bouchiat2010}
and theoretical interest \cite{magarill,ganichev,efetov}.
We note that a case of relativistic dispersion
appears for a flux quantum in long annular Josephson junction,
which has been studied theoretically and experimentally in
\cite{costabile2001,fluxonexp}, but there the dynamics
takes place in 1D while for graphene it is essentially 2D.
In addition, experiments with accelerator beams in crystals
\cite{channeling} show that a crystallographic potential creates
an efficient channeling of relativistic particles.
This gives some indication on  a possible enhancement of ratchet  
transport of electrons in graphene with a periodic array 
of asymmetric antidots.

\begin{figure}
\includegraphics[width=0.48\textwidth]{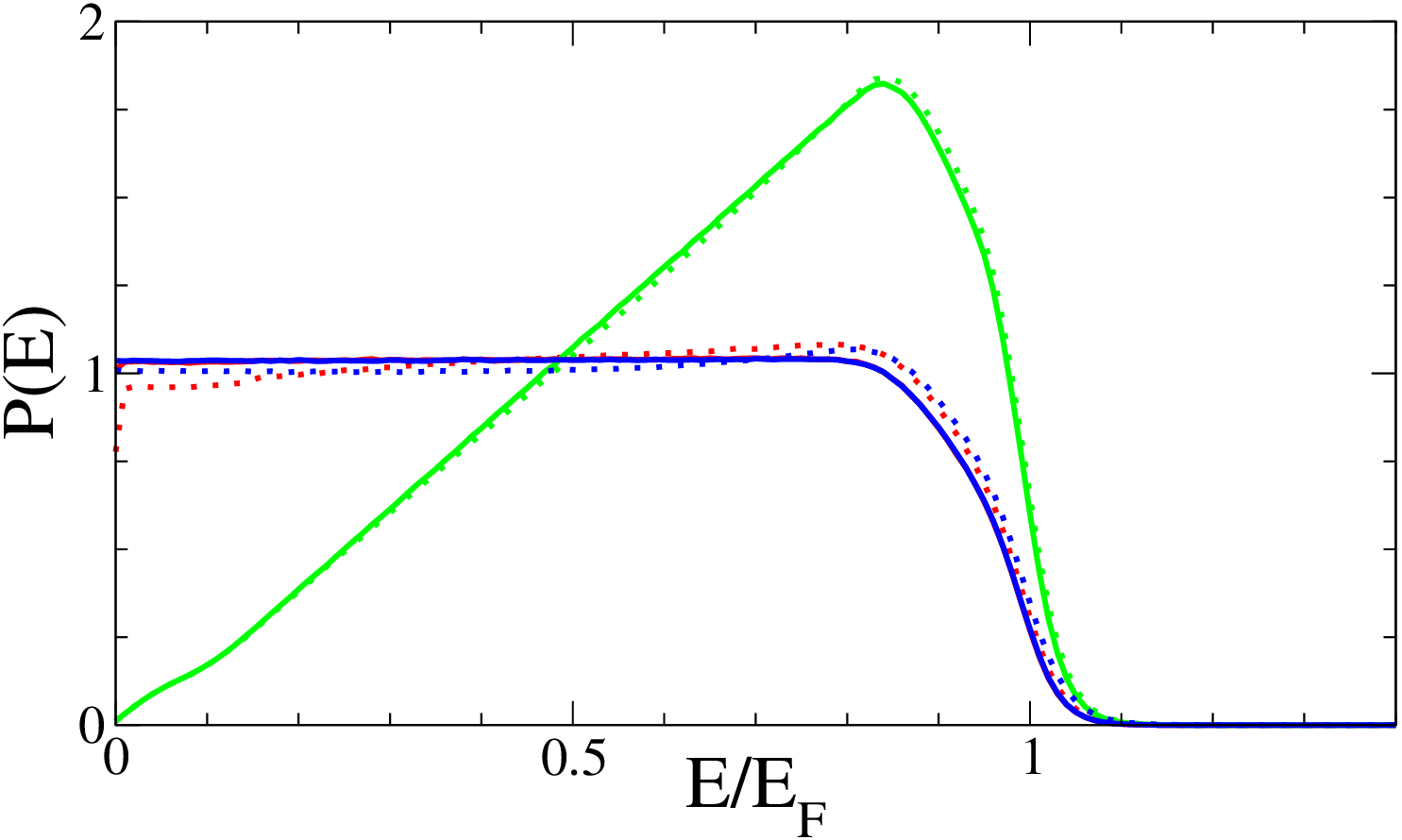}
\vglue -0.2cm
\caption{(Color online) Dimensionless energy distribution functions $P(E)$ obtained
numerically for 2DEG (blue/black curves), model graphene (red/gray curves) 
and graphene (green/gray curves).
Parameters are kept as in Fig.~\ref{fig1} with $t \approx 10^7 r_d/s$ and
temperature fixed to $T/E_F=0.02$ with $fr_d/E_F=0.02$ for full curves, and
$fr_d/E_F=0.1$ for dotted curves (full red and blue curves are overlapped).
}
\label{fig2}
\end{figure}

The dynamics of electron
on a 2D semidisk Galton board, shown in Fig.~\ref{fig1},
is described by the Newton equations 
\begin{eqnarray}
\label{eq2}
d {\mathbf p}/dt = {\mathbf f} \cos \omega t + 
                {\mathbf F_s} + {\mathbf F_i} \; ,   \hskip 2.0cm \\
d {\mathbf r}/dt = s {\mathbf p}/{|\mathbf p|} \; (graphene) \; ; \;\; \nonumber
d {\mathbf r}/dt = {\mathbf p}/m \; (2DEG) \; ,
\end{eqnarray}
where the second equation corresponds to a relativistic case (l.h.s.)
or to a non-relativistic case with an effective mass $m$ (r.h.s.) .
Here the force ${\mathbf F_s}$ describes elastic collisions with semidisks
and ${\mathbf F_i}$ models impurity scattering
on a random angle $\phi_i$ ($|\phi_i| \leq \alpha_i$) 
with an effective scattering time $\tau_i$.

Following \cite{alik2006},
we use the Monte Carlo simulations with the Metropolis algorithm
which keeps noninteracting electrons at 
the Fermi-Dirac distribution with
fixed Fermi energy $E_F$ and temperature $T$. 
As discussed in \cite{alik2006,entin2007,entin2008},
in a wide range,
a variation of energy equilibrium relaxation time $\tau_{rel}$  
does not influence the ratchet velocity ${\mathbf {\bar v}}$.
The later is computed along one or few very long trajectories
with times up to $t = 10^8 r_d/s$.
With this approach we consider three cases
of steady-state distributions. For
2DEG the phase volume is proportional to 
$p dp \sim m d E$, where $E$ is 
an electron energy, hence here the 
probability distribution over energy is 
$P(E) = E_F dW/dE = \rho_{FD}(E)$
where $\rho_{FD}(E) =1/(1+\exp((E-E_F)/T))$
is the Fermi-Dirac distribution at temperature $T$.
For graphene we have the phase volume
$p dp \sim E d E/s^2$
so that the energy distribution has the form
$P(E)=E_F d W/dE = B\rho_{FD}(E)  E/E_F $,
where $B$ is a numerical normalization constant.
We also consider a case of model graphene 
with the energy distribution
 being the same as for 2DEG with
$P(E)= E_F d W/dE = \rho_{FD}(E)$
while the dynamical equations of motion correspond to the
graphene spectrum. Typical examples of the steady-state
distribution $P(E)$ in energy are shown in Fig.~\ref{fig2}
for three models we consider:
graphene, model graphene and 2DEG.

We note that the triangular lattice of disks
had been used already by Galton \cite{galton} to demonstrate 
an emergence of statistical laws in deterministic systems.
According to the mathematical results of Sinai \cite{sinai} the
dynamics is fully chaotic also for semidisk lattice
used here (in absence of impurities, microwave driving and
Metropolis thermostat).

\begin{figure}
\includegraphics[width=0.48\textwidth]{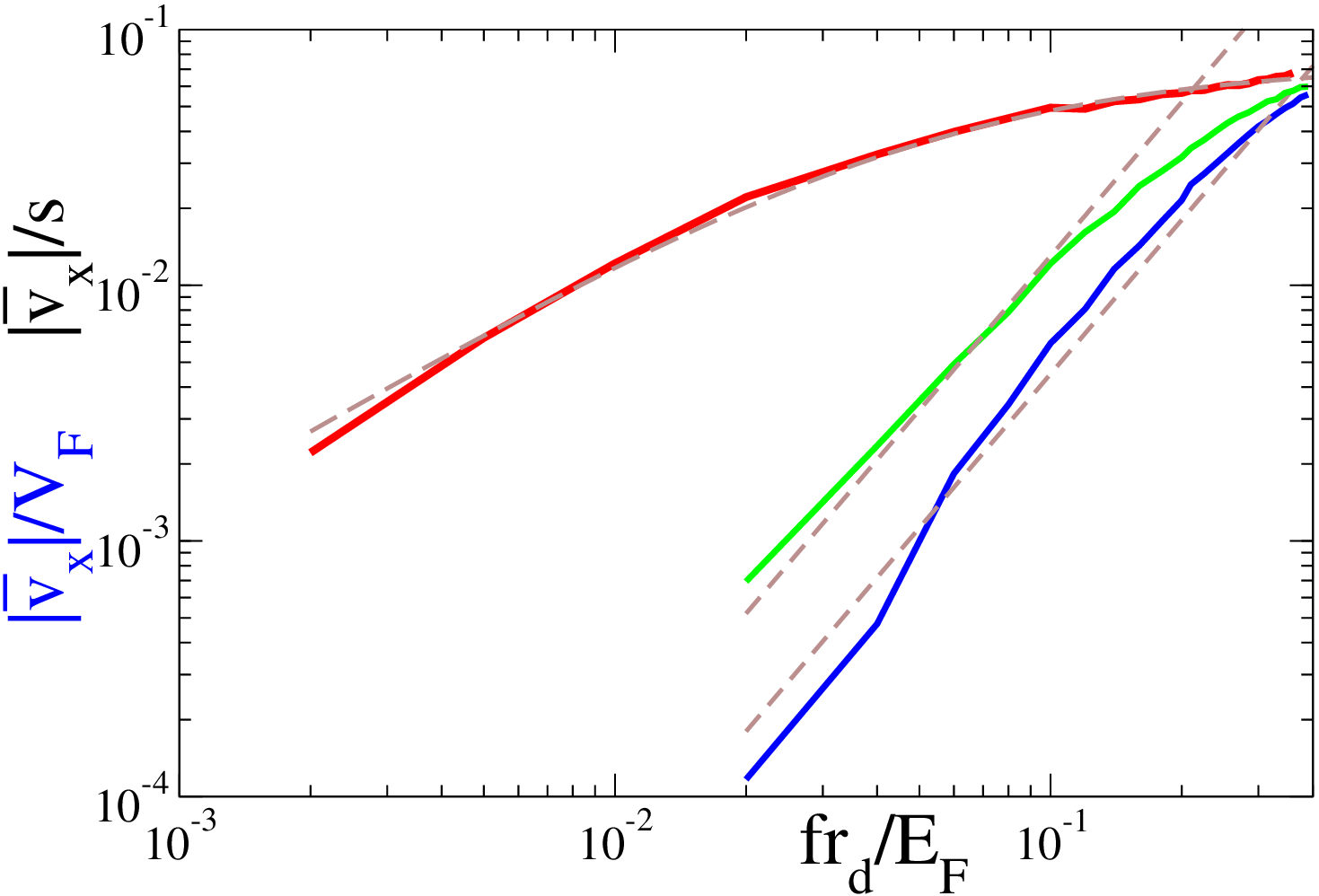}
\vglue -0.2cm
\caption{(Color online) Rescaled ratchet 
velocity $\vert {\bar v}_x \vert$ as a function of rescaled 
force $f r_d/E_F$ at polarization $\theta=0$
for graphene (green/gray curve), model graphene 
(red/gray curve) and 2DEG (blue/black curve);
when $f$ is changing the system parameters are kept
as Fig.~\ref{fig1}. 
The bottom straight dashed line shows the fit dependence 
for 2DEG (\ref{eq3}), the top dashed curve shows the fit dependence
for model graphene (\ref{eq4}) and 
the middle straight dashed line shows the fit dependence 
for graphene (\ref{eq5}).
}
\label{fig3}
\end{figure}

A typical example of trajectory for graphene 
is shown in Fig.~\ref{fig1}. The dynamics is clearly chaotic 
on one cell scale, while on large scale it shows diffusion and
ratchet transport directed along $x$-axis. The ratchet displacement
is significantly larger for model graphene compared to usual 2DEG
with approximately the same parameters. Even more striking
a decrease of the driving force by a factor $4$ 
gives a significantly smaller reduction of
the ratchet displacement compared to factor $16$
expected from the theory for 2DEG (\ref{eq1}).  
Thus for the model graphene the relativistic graphene ratchet has a strong 
enhancement compared to the usual 2DEG case studied before.
The case of graphene has more strong ratchet compared
to 2DEG but it is less strong compared to the model graphene.

Detailed analysis of this enhancement 
and comparison of ratchet transport for graphene, model graphene and 2DEG,
as a function of microwave driving force,
are  shown in Fig.~\ref{fig3}. 
We fix $V_F/s=2$ choosing $p_F$ to be the same for graphene and 2DEG
that corresponds to the same electron density $n_e$.
For 2DEG the velocity of ratchet
drops quadratically with force and is well described by the dependence
\begin{equation}
\label{eq3}
\vert {\bar v}_x \vert/V_F = C_F (f r_d/E_F)^2 \; , \; \; C_F =0.45 \;\; ,
\end{equation}
thus being in a good agreement with numerical data and theory
presented in \cite{alik2006,entin2007,entin2008}. 
In a case of model graphene the field dependence is strikingly different
and can be approximately described by the equation
\begin{equation}
\label{eq4}
\vert {\bar v}_x \vert/s = C_{g1} f r_d/(E_F + C_{g2} f r_d) \; , 
\;  C_{g1} =1.39, C_{g2}=18.87  .
\end{equation}
According to Fig.~\ref{fig3} and Eqs.~(\ref{eq3}),(\ref{eq4})
we have the enhancement factor for model graphene of approximately
$100$ and  $10^3$ at $f r_d/E_F=0.02$ and $0.002$ respectively.

However, for graphene we find approximately quadratic field dependence with
\begin{equation}
\label{eq5}
\vert {\bar v}_x \vert/s = C_g (f r_d/E_F)^2 \; , \; \; C_g =1.3 \;\; ,
\end{equation}
formal fit gives a power exponent $1.8$ being close to $2$.
Thus in this case the ratchet velocity is comparable with the one
of 2DEG.

\begin{figure}
\includegraphics[width=0.48\textwidth]{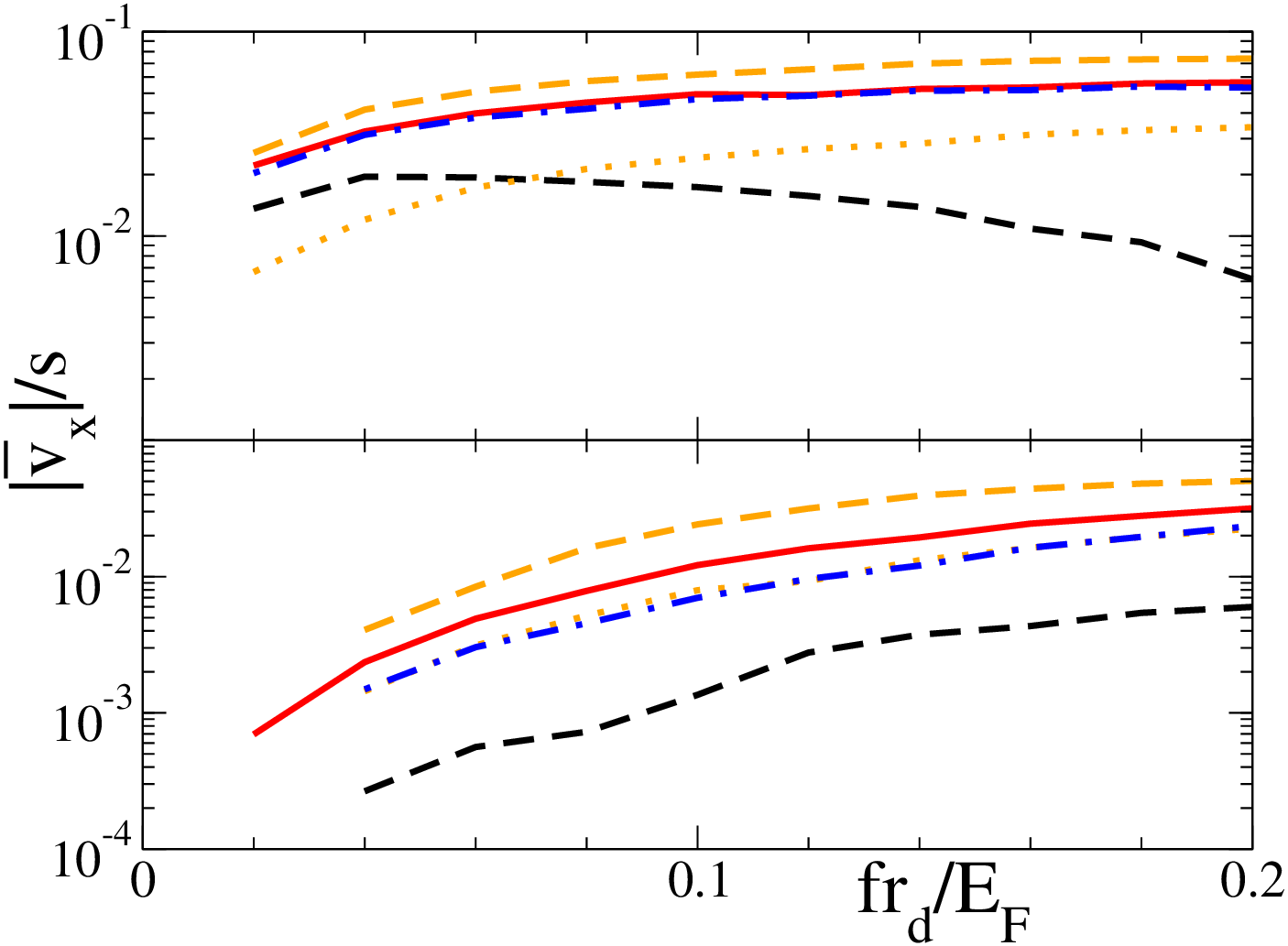}
\vglue -0.2cm
\caption{(Color online) Rescaled ratchet 
velocity $\vert {\bar v}_x \vert$ as a function of rescaled 
force $f r_d/E_F$ at polarization $\theta=0$
for model graphene in top panel, and graphene in bottom panel,
for the same values given in Fig.~\ref{fig1} with modifications of 
frequency, temperature or impurity scattering angle.
Dashed orange/gray curves are for
$\omega r_d/s=0.25$, $T/E_F=0.05$, $\alpha_i=\pi/10$;
solid red/gray curves are for
$ \omega r_d/s=0.5$, $T/E_F=0.05$,  
$\alpha_i=\pi/10$ (Fig. \ref{fig3} case);
dotted-dashed blue/black line are for
$ \omega r_d/s=0.5$, $T/E_F=0.3$,  
$\alpha_i=\pi/10$;
dotted orange/gray curves are for 
$ \omega r_d/s=0.5$, $T/E_F=0.05$,  
$\alpha_i=\pi$;
dashed black curves are for
$\omega r_d/s=0.1$, $T/E_F=0.05$, $\alpha_i=\pi/10$.
}
\label{fig4}
\end{figure}

Data presented for model graphene in the top panel of Fig.~\ref{fig4}
show that the ratchet velocity is only weakly affected by
increase of temperature $T$, which can become comparable with $E_F$,
if the rate of impurity scattering is kept fixed.
This is in agreement with the known results for 2DEG 
\cite{alik2006,entin2007}. An increase of impurity
scattering (increase of $\alpha_i$) gives a reduction
of ${\bar v}_x$ but still the dependence on
$f$ remains approximately linear for $fr_d<<E_F$. A decrease of frequency
gives only a slight increase of ${\bar v}_x$ for
$\omega r_d/s \leq 0.5$ while for $\omega r_d/s \geq 1$
we start to see a drop of ${\bar v}_x$ with $\omega$.
Such a dependence is 
in agreement with general theory
\cite{entin2007} according to which 
the ratchet velocity is independent 
of frequency for $\omega r_d/s \ll 1$
and drops with frequency for  $\omega r_d/s \gg 1$.
It is interesting to note that
at $\omega r_d/s =1$ the velocity ${\bar v}_x$
starts to decrease at large $f$. We will discuss this point later.
For the graphene case the dependence of rescaled force $f$ is shown in 
the bottom panel of Fig.~\ref{fig4} at same parameters.
Here for all cases the velocity of ratchet increases with $f$,
it is smaller compared to the case of model graphene in the top panel.

\begin{figure}
\includegraphics[width=0.48\textwidth]{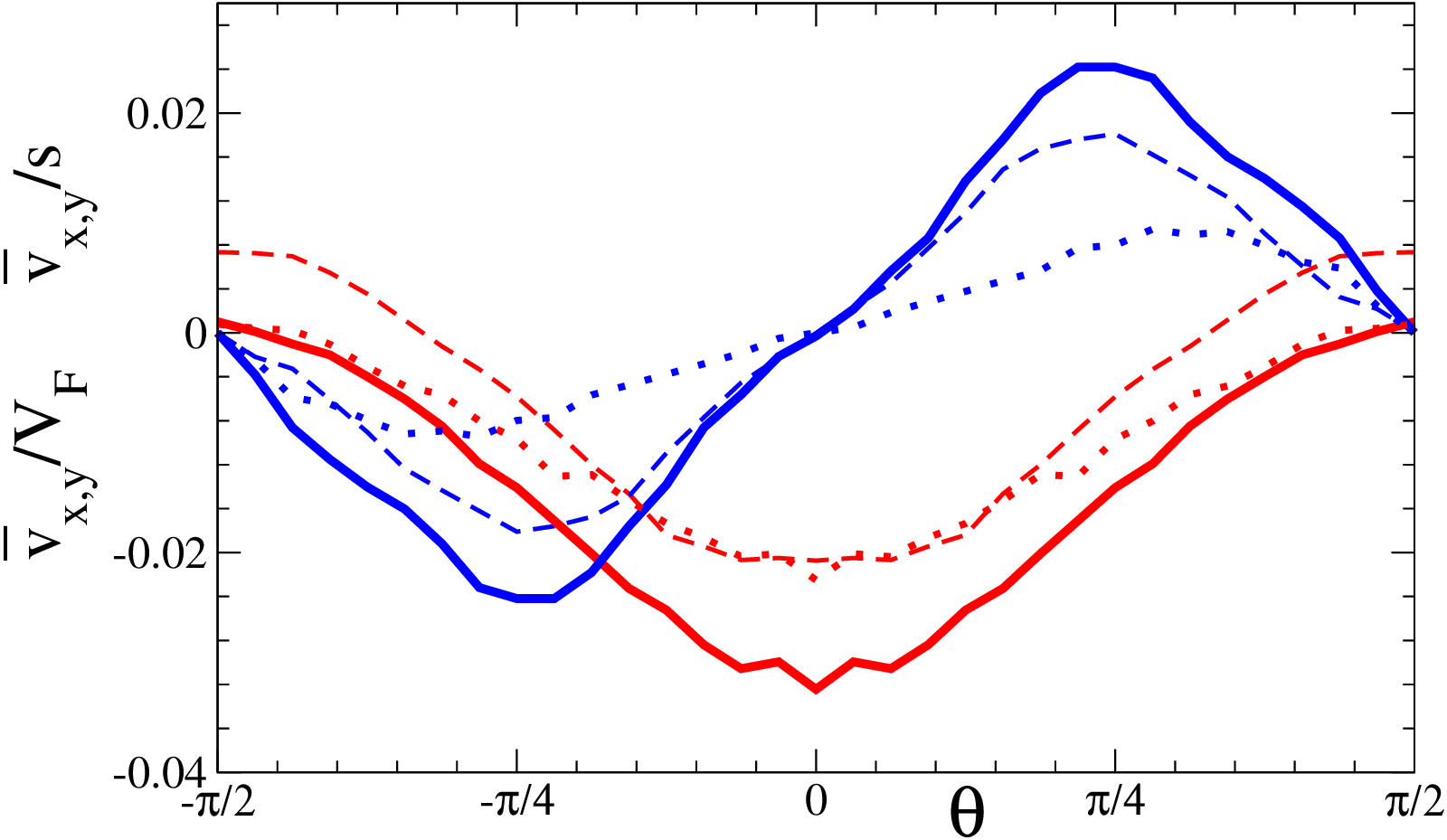}
\vglue -0.2cm
\caption{(Color online)  Polarization dependence of ratchet velocity 
in $x$ (red/gray curves) and $y$ (blue/black curves) directions.
Data for graphene are taken at $f r_d/E_F=0.2$ and
$T/E_F=0.02$ (full curves), $0.3$ (dotted curves).
Data for 2DEG are shown for $f r_d/E_F=0.2$ and
$T/E_F=0.02$ (dashed curves). Other parameters are as Fig.~\ref{fig1}.
}
\label{fig5}
\end{figure}

\begin{figure}
\includegraphics[width=0.48\textwidth]{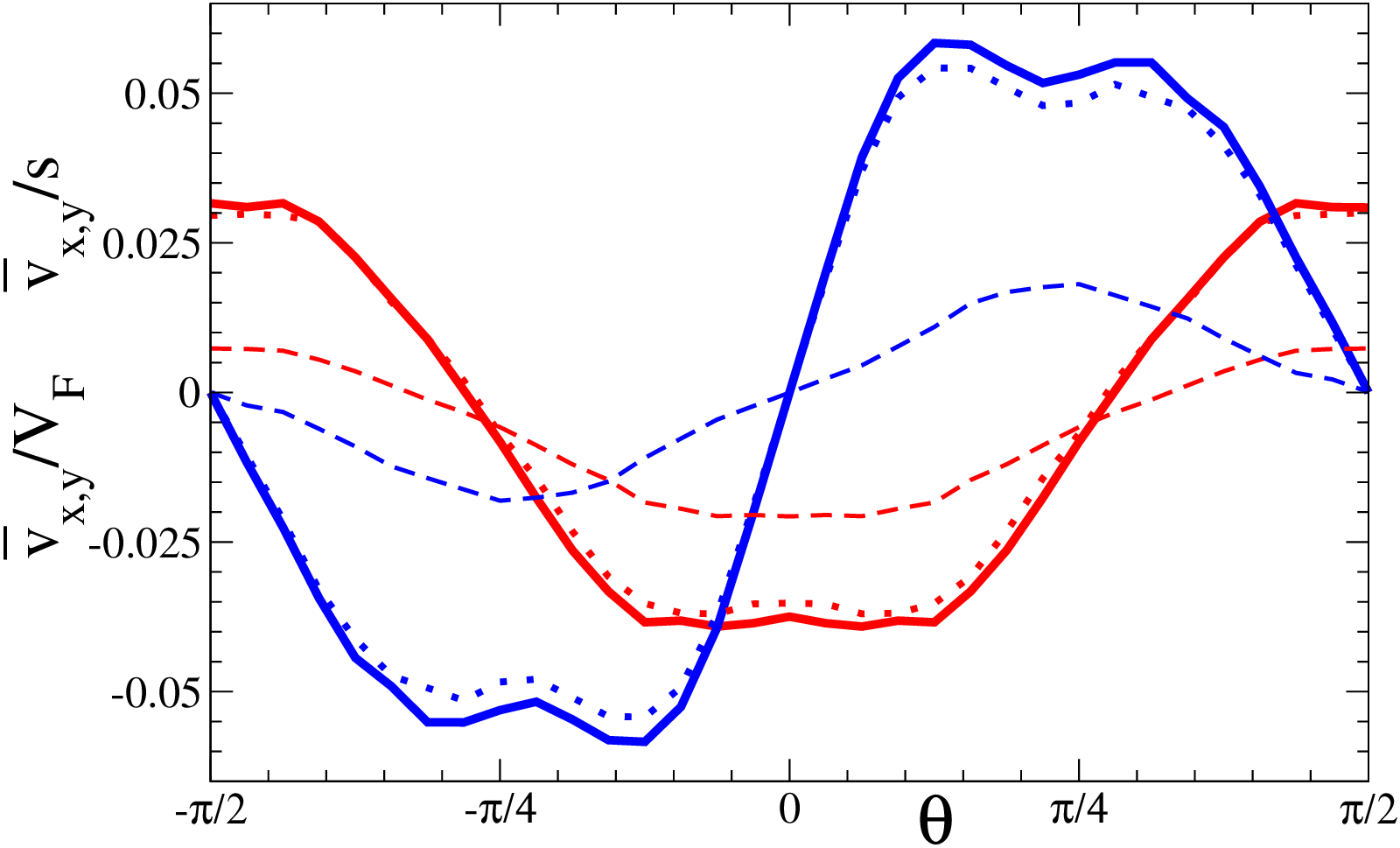}
\vglue -0.2cm
\caption{(Color online) Polarization dependence of ratchet velocity
in $x$ (red/gray curves) and $y$ (blue/black curves) directions.
Data for model graphene are taken at $f r_d/E_F=0.05$ and
$T/E_F=0.02$ (full curves), $0.3$ (dotted curves).
Data for 2DEG are shown for $f r_d/E_F=0.2$ and
$T/E_F=0.02$ (dashed curves). Other parameters are as Fig.~\ref{fig1}.
}
\label{fig6}
\end{figure}

\begin{figure}
\includegraphics[width=0.235\textwidth]{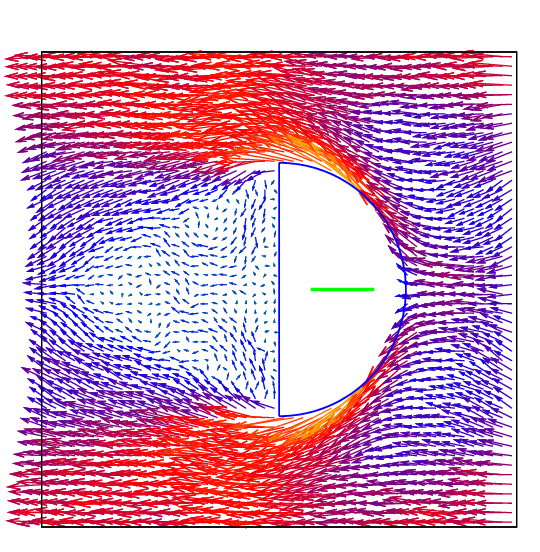}
\includegraphics[width=0.235\textwidth]{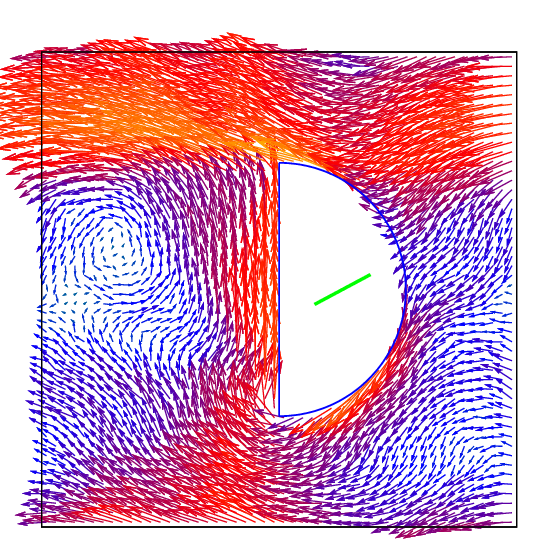}\\
\includegraphics[width=0.235\textwidth]{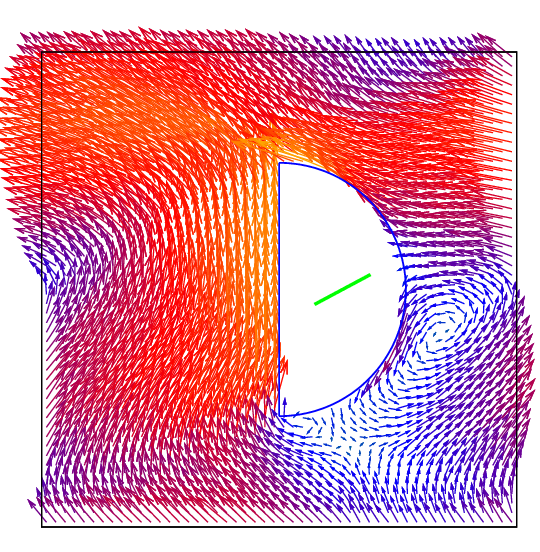}
\includegraphics[width=0.235\textwidth]{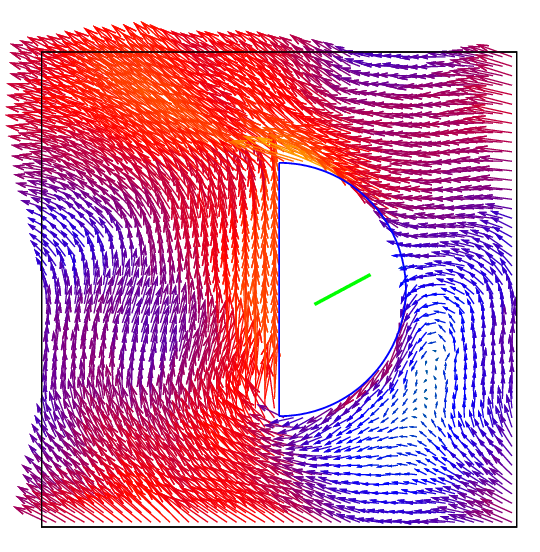}
\caption{(Color online) 
Map of local averaged velocities inside one cell in plane $(x,y)$
for graphene at $\theta=0$ (top left panel) and
$\theta=5\pi/32$ (top right panel) with parameters of Fig.~\ref{fig5};
for model graphene at
$\theta=5\pi/32$ (bottom left panel) and parameters of
Fig.~\ref{fig6};
for 2DEG at $\theta=5\pi/32$ (bottom right panel)
and parameters of
Fig.~\ref{fig5};
polarization is shown by bar inside semidisk.
The velocities are shown by arrows which size is proportional 
to the velocity amplitude,
which is also indicated by color (from large (yellow/gray) 
to small (blue/black) amplitudes).
}\label{fig7}
\end{figure}

The polarization dependence of ratchet flow is shown in Fig.~\ref{fig5}
for 2DEG and graphene; for 2DEG and model graphene the polarization 
dependence is shown in Fig.~\ref{fig6}. For 2DEG the dependence is close
to those of Eq.~(\ref{eq1}) being in agreement with previous studies 
\cite{alik2006,entin2007,entin2008}. 
A similar polarization dependence is present also
for graphene.
In contrast to that
for model graphene the dependence on $\theta$ is not purely harmonic
and appearance of flat domains near velocity maxima
is well visible. In part the origin of this flattering
can be understood from the picture of average flow inside 
one periodic cell shown in Fig.~\ref{fig7}. While for
$\theta=0$ the flow is relatively regular, for
$\theta=5\pi/32$ there is emergence of vortexes
that may be at the origin of flattering in Fig.~\ref{fig6}.
It is clear that a theoretical description of such a vortex flow
is a challenging task for further studies.
We note that in agreement with theory \cite{entin2007}
the polarization dependence  on temperature is rather weak.
It is interesting that in the case of graphene even unpolarized
radiation gives an average current along $x-$direction
with ${\bar v_x} <0$. This is different from 2DEG case
where for semidisks there is no directed current
for unpolarized radiation \cite{entin2007}.
We attribute this to the fact that for graphene 
$P(E)$ becomes energy dependent that
acts in a similar way when scattering time $\tau(E)$
is energy dependent and when even unpolarized
radiation creates a directed transport \cite{entin2007}. 

The important theoretical task is to understand the origin
of a slow decrease of ratchet velocity with $f$ for model graphene
while for the graphene case the dependence remains quadratic as for 2DEG. 
We argue that this is due to the linear dispersion law for graphene which 
drastically modify the dependence of velocity 
${\mathbf  v}$ on momentum and force.
Indeed, in absence of collisions and impurities we have
${\mathbf  v} = ({\mathbf  p_0} + {\mathbf f}/\omega \sin \omega t)/m$
where ${\mathbf  p_0}$ is initial momentum. Thus a small force
gives only a small oscillating component of velocity and 
thus the ratchet flow appears only in a second order
perturbation theory being proportional only to $f^2$ \cite{entin2007}.
The situation is strongly different for model graphene.
In absence of collisions and impurities we obtain from (\ref{eq1})
${\mathbf  v} = s 
({\mathbf  p_0} + {\mathbf f}/\omega \sin \omega t)/|({\mathbf  p_0} + 
{\mathbf f}/\omega \sin \omega t)|$. Thus even for small force we have
a large  variation of velocity direction being of the order of radian
for $|{\mathbf  p_0}| \sim  f/\omega$. These direction variations have
many frequency harmonics in contrast to 2DEG. In presence of semidisk
asymmetry and relaxation such oscillations should create
a directed transport with ratchet velocity ${\bar v} \sim s$.
However, in the Fermi-Dirac distribution 
the fraction $w$ of electrons with such small momenta $p_0 < f/\omega$ is  
$w \sim f s/(\omega E_F) \sim f r_d/E_F$, where we took into account that
in the case of small $\omega$ the collisions with semidisks 
will restrict the length $s/\omega$ by a length
proportional to $r_d$. These arguments lead to the ratchet velocity
${\bar v} \sim s w \sim s f r_d/E_F$ being compatible with
the dependence (\ref{eq4}) found numerically. According to them
a physical origin of slow decrease of ${\bar v}$ with $f$
is related to linear dispersion and Dirac singularity in graphene.
In such a case electrons with  momentum $p \ll p_F$
still have a large velocity $s$ and give a large contribution to
the ratchet flow in contract to 2DEG.

It is interesting to note that at high frequencies $\omega$
only few collisions happen during a period of 
oscillations so that in  average we have 
$<|{\mathbf  p}|> \approx \sqrt{p_0^2+f^2/2\omega^2}$
that looks like an appearance of effective mass
at large $f$. This mass effectively drives the system to a situation
similar to 2DEG given a reduction of ${\bar v}$
at large $f$ (see black dashed curve in Fig.~\ref{fig2} inset).

The above arguments attribute the strong ratchet transport to
contribution of trajectories in a vicinity of Dirac critical point
in the model graphene. In this model the measure $P(E)$ of such trajectories
is independent of energy $E \rightarrow 0$.
However, for real graphene this measure drops with energy 
$P(E) \propto E$ and therefore the contribution of this region
gives a weaker quadratic dependence on the driving force $f$
that is compatible with the numerically established dependence
(\ref{eq5}). The numerical factor $C_g$ is larger than in 2DEG
but the enhancement of ratchet transport is not so strong as for model graphene.

The electron current produced by ratchet flow
is given by relation $j =e n_e {\bar v}$. According to (\ref{eq3}),(\ref{eq5})
we have for 2DEG
\begin{eqnarray}
\label{eq6}
j_F \approx 0.4 e n_e V_F (f r_d/E_F)^2 \hskip 2.0cm \\ \nonumber
\approx J_F  (n_0/n_e)^{1/2} (f r_d/1 K)^2 \; , \; J_F \approx 4 \cdot 10^{-5} A/cm \; ;
\end{eqnarray}
and for graphene 
\begin{eqnarray}
\label{eq7}
j_g \approx  1.3 e n_e s (f r_d/E_F)^2 \approx 1.3 e (f r_d)^2/\pi s \hbar^2 \\ \nonumber 
\approx J_g  (f r_d/1 K)^2  \; , \; J_g \approx 1.5 \cdot 10^{-5} A/cm \; ;
\end{eqnarray}
where we normalize data in respect to typical parameters
of 2DEG in \cite{portal2008} with  $n_0 = 2.5 \cdot 10^{11} cm^{-2}$, $m=0.067m_e$,
$f/e= 1 V/cm$, $r_d=1 \mu m$, $f r_d \approx 1K$, $E_F \approx 100 K$,
$V_F \approx 2.2 \cdot 10^7 cm/s$
and usual parameters of graphene
$s \approx 10^8cm/s$
\cite{geimrmp2009,stormer2008}.
It is interesting that for graphene the ratchet current is independent of
electron density $n_e$ (we use the relation
$n_e=p_F^2/\pi \hbar^2$  \cite{geimrmp2009}).

The comparison of estimates for 2DEG (\ref{eq6}) and graphene (\ref{eq7})
show that the ratchet currents in these two materials
have comparable values.
 Of course, the above 
Eqs.~(\ref{eq6}),(\ref{eq7}) assume that 
the mean free path in graphene is larger
than the semidisk radius $r_d \sim 1 \mu m$.
For 2DEG the mobility drops significantly with increase of temperature
and the ratchet transport disappears according to theory
at small mean free path \cite{entin2007} and according to
experimental results presented in \cite{portal2008}.
For 2DEG the experiments \cite{portal2008}
show that the ratchet transport at such $r_d$
values persists only up to $T \approx 70 K$
while at room temperature the mean free path becomes smaller than 
semidisk size and the ratchet effect disappears.
According to experiments with suspended monolayer graphene
\cite{stormer2008,andrei2008} 
the mean free path can be larger
than $r_d \sim 1  \mu m$ at room temperature
so that the graphene ratchet transport
should be well visible at room temperature. At present it is possible to 
realize large size samples with epitaxial graphene \cite{berger1,berger2}
and chemical vapor deposition \cite{colombo2010}
with room temperature mobility of $20000 cm^2/Vs$. 
For graphene even at room temperature the mean free path can be rather large
and thus the ratchet transport can be significant even at room temperature.
For strong microwave fields of $10 V/cm$ ratchet currents can be of the order 
of $10^{-3} A/cm$.

In conclusion, our theoretical studies show that 
graphene ratchet transport in asymmetric antidot arrays, 
at micron antidot size,
should be well visible at room temperature in contrast to the 
case of 2DEG. This shows that such graphene structures can have future promising 
applications for simple room temperature sensors of microwave and terahertz radiation.
A decrease of antidot size can make such structures to be 
sensitive to infrared radiation with possible photovoltaic applications.

We thank A.D.Chepelianskii for useful discussions and for pointing to us
promising properties of graphene \cite{geimmobil,stormer2008,andrei2008} 
for ratchet transport. We also thank M.V.Entin and L.I.Magarill
for critical remarks.
This work is supported in part by ANR PNANO project NANOTERRA.

\end{document}